# CRIMINAL LIABILITY IN AI-ENABLED AUTONOMOUS VEHICLES: A COMPARATIVE STUDY

*Dr.Manjit Singh[*]*
*Sahibpreet Singh[**]*

## ABSTRACT

*AI is revolutionizing transportation by making it more sustainable. This application in autonomous vehicles has its own set of complexities concerning liability in case of infractions. The methodology employed in this study involves a comparative legal analysis approach. This includes a comprehensive analysis of primary legal documents to understand the current legal landscape in the selected jurisdictions. Additionally, the study draws on a real-world comparative analysis and examines liability claims to gain practical insights into the legal complexities. Secondary sources include academic literature, industry reports and news articles. This paper examines various aspects of criminal responsibility of AI-based AVs, drawing comparisons among US, Germany, UK, China and India. The rationale for comparing these countries lies in their diverse legal frameworks. These countries were chosen for their technological advancements and contrasting regulatory approaches to liability in AI-enabled autonomous vehicles. The goal is to compare how different countries have approached this problem by analyzing their legal frameworks and responses to it. It explores various approaches for ascertaining human errors that result in crime, such as intervention or moral agency on the part of AI, and identification of the primary offenders in incidents involving AVs. However, it shows that every country has its own unique way within its respective jurisdiction. For instance, India and USA have a loose interweaving network of state laws, while UK made a pioneering piece of legislation in 2018 called the Automated and Electric Vehicles Act, 2018. Germany applies strict safety standards and distinguishes liability based on the operating mode of the vehicle. Contrarily, China also aims to establish a very strict liability regime for AVs. Lastly, as an outcome of this study, it was found that there is a pressing need for globally agreed upon legal standards to encourage technological advancements, ensuring there is innovation invoking minimum risk.*

---

[*]Assistant Professor, Department of Laws, Guru Nanak Dev University, Amritsar, Punjab.
[**]LLM (2023-24), Department of Laws, Guru Nanak Dev University, Amritsar, Punjab.

## 1. INTRODUCTION

Self driving cars are the kind of vehicles that employ artificial intelligence as well as other sensors to operate autonomously. This is done without human intervention or supervision. These vehicles can make transportation safer and more efficient by helping people get where they are going easier, in an environmentally sustainable manner.[1] AVs[2] also have significant legal and ethical connotations, especially criminal liability, since they pose questions about who is at fault for their actions and outcomes. However, the shift from conventional vehicles to AVs is neither consistent nor continues in a straight line; rather, it progresses in degrees with different levels of automation, capabilities and applications. The SAE[3] has come up with six levels of automation for on-road vehicles.[4] They are as follows:

**Level 0: No Automation-** The human driver performs all driving tasks.

**Level 1: Driver Assistance-** It is specified that the vehicle system shall be able to provide assistance to the human operator with either steering or acceleration/deceleration. Nonetheless, it is clearly stated that these two cannot be offered simultaneously.

**Level 2: Partial Automation-** The automotive system assists the human driver in steering and acceleration/deceleration. However, the driver must always be ready to take control of the vehicle.

**Level 3: Conditional Automation-** The vehicle can drive by itself under some conditions. However, a human being must be ready to take over control at any moment, should there be a need for assistance from him/her or a fault happens in the system.

**Level 4: Elevated Degree of Automation-** These are systems that perform all driving tasks under specific conditions without requiring participation from humans or supervision by drivers, like in a traditional car.

**Level 5: Complete Automation-** The system performs all driving tasks under all circumstances. In no way does it involve a person who is expected to participate in whatever form.

Indeed, the issue of criminal liability is a significant and contentious aspect within the legal framework for AVs or their AI[5] systems. Criminal liability pertains to the legal obligation incurred for the commission of a crime, which could lead to sanctions. It includes fine, imprisonment or other penalties.[6] The primary elements consist of the two:

i. Actus reus (the wrongful act) and
ii. Mens rea (the guilty mind).[7]

---

[1] Inclusion Cloud, "The Future of Autonomous Vehicles: Evolution, Benefits, and Challenges" *available at*: https://inclusioncloud.com/insights/blog/future-autonomous-vehicles/ (last visited November 7, 2023).
[2] Autonomous Vehicles.
[3] Society of Automotive Engineers.
[4] SAE International, "Taxonomy and Definitions for Terms Related to Driving Automation Systems for On-Road Motor Vehicles J3016_201806," 2023 *available at*: https://www.sae.org/standards/content/j3016_201806/ (last visited November 7, 2023).
[5] Artificial Intelligence.

However, the process of integrating these principles into the operations is neither direct nor unequivocal. The inherent complexities of these advanced technologies pose significant challenges to the unambiguous incorporation of existing legal elements. According to some scholars, there are three basic models to cope with this phenomenon within the current definitions of criminal law.[8] These models are:

a. **The Perpetration-by-Another Liability Model:** In this model, a human being who uses, controls or directs the AI system that commits a crime will be held liable as the perpetrator, whereas the AI system is seen as an instrumental tool.

b. **The Natural Probable Consequence Liability Model:** This model holds that if someone creates, programs or allows an autonomous AI system to commit a crime, then they are liable as an accessory since it was foreseeable that such criminal conduct would naturally flow from their acts or omissions.

c. **The Direct Liability Model:** This model holds that where an AI system can be said to have had enough autonomy, intelligence and moral agency, it may also be considered criminally responsible as a separate legal person.

These models have different benefits and drawbacks and thus, they may be used differently depending on the degree of automation, capacity as well as purpose for which the AI system is put in place. Some legal systems could blend various models or even create others to respond specifically to challenges posed by AI systems.

## 2. CRIMINAL LIABILITY OF SDVs

A primary obstacle in the regulation of these vehicles is the identification of the party to be held legally accountable upon any violation.[9] The question of who bears legal responsibility for acts done by AV remains unanswered. Liability could be attributed to either humans operating them, AI system running them, vehicle manufacturers, software developers making programs that make these cars function, service provider or a combination of these parties. This issue can be approached differently in different jurisdictions depending on their underlying principles, values and objectives. Various legal viewpoints on the criminal liability of autonomous vehicles within the jurisdiction of India and other nations are discussed hereunder.[10]

---

[6] Andrew Ashworth and Jeremy Horder, *Principles of Criminal Law*, 7th edition (Oxford University Press, Oxford, 2013).
[7] David Ormerod et al., *Smith and Hogan's Criminal Law* (Oxford University Press, 2015).
[8] Prof. Gabriel Hallevy, "The Basic Models of Criminal Liability of AI Systems and Outer Circles" *SSRN Electronic Journal* (2019).
[9] Ziya Altunyaldiz, *Legal Aspects of "Autonomous" Vehicles*, 2020.

### 2.1 INDIA

India is a common law jurisdiction following the principles of tort law. It includes the concept of negligence, strict liability and product liability as well. India currently lacks any specific legislation addressing autonomous vehicles. Motor Vehicles Act, 1988 is the main ruling that governs road transport & traffic in India. The MVA[11] defines a motor vehicle as

> "any mechanically propelled vehicle adapted for use upon roads whether the power of propulsion is transmitted thereto from an external or internal source".[12]

The act mandates that every motor vehicle must be registered[13], insured[14] and driven by an individual holding a valid driving licence.[15] MVA outlines various offences and penalties for contravention of its provisions or inflicting harm on others through the use of a motor vehicle.[16] It also provides for offences and penalties where these provisions are contravened or harm is caused to others using the motor vehicle. However, this has nothing to do with self-driving cars or their criminal liability as provided for under the MVA. In addition, it should be noted that the MVA assumes that there is always a human driver behind a wheel whose actions can be attributed to him; it does not consider possibilities such as transference of control or accountability to an artificial intelligence system or even another entity. Thus, if an accident occurs or a traffic law is broken by a self-driving car, under the MVA, the human driver is more likely to be assumed responsible unless they can prove otherwise.[17] This can lead to unfair and absurd circumstances for the human driver, especially when he did not know or participate in the failure or mistake of that self-driving car.[18]

---

[10] Dhawal Srivastava, "Legal issues related to autonomous vehicles" *available at*: https://blog.ipleaders.in/legal-issues-related-autonomous-vehicles/ (last visited November 7, 2023).
[11] The Motor Vehicles Act, 1988 (Act No. 59 of 1988).
[12] The Motor Vehicles Act, 1988 (Act No. 59 of 1988), s. 2(28).
[13] The Motor Vehicles Act, 1988 (Act No. 59 of 1988), s. 39.
[14] The Motor Vehicles Act, 1988 (Act No. 59 of 1988), s. 146.
[15] The Motor Vehicles Act, 1988 (Act No. 59 of 1988), s. 3.
[16] The Motor Vehicles Act, 1988 (Act No. 59 of 1988), Chapter X and XI.
[17] Essenese Obhan and Shourya Paul, "Driving into the Future: Regulating Autonomous Vehicles" *available at*: https://www.mondaq.com/india/patent/1239240/driving-into-the-future-regulating-autonomous-vehicles (last visited November 7, 2023).
[18] Yineng Xiao and Zhao Liu, "Accident Liability Determination of Autonomous Driving Systems Based on Artificial Intelligence Technology and Its Impact on Public Mental Health," 2022 *Journal of Environmental and Public Health* e2671968 (2022).

Consider the case in which a self-driving car is given an erroneous software update[19] or it is hacked by a third party[20] and ends up harming others. In this case, the driver may face difficulty trying to prove his innocence or lack of fault.[21] Besides, this may also discourage adoption and innovation of autonomous vehicles in India, where potential users fear legal consequences and liabilities. Thus, it becomes imperative for India to formulate a specific regulatory framework to comprehensively address the complexities surrounding autonomous vehicles and their potential criminal liability. The following legislations could have potential connotations for SDVs[22] as well:

   i. Consumer Protection Act is an enactment to advocate for the rights of consumers. This legislation ensures end user safety in relation to defective commodities. It is also applicable in cases involving unsatisfactory service and inequitable trade practices.[23] This act aims to protect consumers' interests by establishing authorities for quick administration and settlement of consumers' grievances.[24] In addition, it includes certain provisions relating to product liability, misleading advertisements and e-business.[25] The manufacturers might have to be held responsible if these cars cause any harm due to any malfunction. The service providers might also be held responsible in one case or another.

   ii. IT Act, 2000[26] confers legal recognition to transactions carried out through electronic data interchange. It may involve other means of electronic communication as well.[27] This contains provisions relating to cyber crimes and their penalties, as well as rules for the use of digital signatures and electronic records. There is also the provision of setting up a Cyber Appellate Tribunal.[28] Similarly, in relation to autonomous cars, it can be said that this might cover software up-dates; digital transactions and privacy; and digital transaction issues related to self driving vehicles. For instance, where an autonomic car receives a defective software up-date or is hacked by a third party, leading to injuries to others, then the situation may fall under the purview of this law.[29]

---

[19] BBC News, "Tesla withdraws self-driving beta over software issues" *available at*: https://www.bbc.com/news/technology-59037344 (last visited November 7, 2023).
[20] Sascha Brodsky, "How Self Driving Cars Can Be Hacked" *available at*: https://www.lifewire.com/how-self-driving-cars-can-be-hacked-5114337 (last visited November 7, 2023).
[21] Rohit Ray, "Liability For Self Driving Vehicles: Is There Anyone To Blame?" *available at*: https://www.livelaw.in/lawschoolcolumn/liability-for-self-driving-vehicles-automated-cars-driverless-technology-212778 (last visited November 7, 2023).
[22] Self Driving Vehicles.
[23] BYJUS, "Consumer Protection Act, 2019 - A Brief Overview" *available at*: https://byjus.com/free-ias-prep/consumer-protection-act-2019/ (last visited November 7, 2023).
[24] Ministry of Consumer Affairs, *FAQs on Consumer Protection Act, 2019*, 11 September 2023.
[25] Citizen Consumer and Civic Action Group, *Salient Features of the Consumer Protection Act, 2019*, October 2019.
[26] The Information Technology Act, 2000 (Act No. 21 of 2000).
[27] Dr. Harisingh Gour Vishwavidyalaya Sagar, "The Information Technology Act, 2000" *available at*: https://dhsgsu.edu.in/images/Reading-Material/Law/UNIT-IV-Second.pdf (last visited November 7, 2023).
[28] Team ZCySec, "Key Provisions of the Information Technology (IT) Act, 2000" *available at*: https://zcybersecurity.com/key-provisions-of-the-information-technology-it-act-2000/ (last visited November 7, 2023).

However, it is vital that we recognize that, though these Acts have some consumer protection measures as well as regulations on aspects of IT[30], but they don't specifically deal with the unique problems posed by self-driving cars. It is therefore necessary for India to come up with special legislation or regulations on self-driving cars' criminal liability. This would assist in diverging such matters like accountability during malfunctions or accidents and other things like data privacy and security, maintenance, together with software updates.

### 2.2 UNITED KINGDOM

AEVA[31] has been passed by the UK. It defines an automated vehicle as a motor vehicle which is designed or adapted to be capable of safely driving itself without having to be monitored by an individual.[32] According to the AEVA, whenever an automated vehicle is operating on its own, the insurer of such a vehicle will have the responsibility of compensating for any injury caused to another person or property.[33] However, the insurer can disclaim or limit its liability provided that the accident was triggered by unauthorized modification of such a vehicle or where its software has not been updated when required.[34] The amount expended may be claimed from those responsible for making the alteration without authorization or failing to update it.[35] Additionally, AEVA enables the Secretary of State to make regulations for testing and certifying self-driving cars.[36]

### 2.3 GERMANY

The Road Traffic Act (StVG) of Germany has been changed to allow for the use of highly or fully automated driving functions. Such a system is described as one that can take over the driving task within a certain operational area, although the driver can switch off such systems at any time.[37] The StVG[38] still requires the driver to be responsible for any destruction caused by the vehicle if he is not guilty.[39] Nonetheless, where this damage arises out of defects in an automatic drive system, it becomes the responsibility of either the car or the system manufacturers limited to a maximum amount of 10 million Euros.[40] Also, in accordance with StVG requirements, the driver must remain alert, be ready to regain control of the vehicle at any given moment and adhere to all orders given by the automated driving function.[41] Additionally, StVG also asks that there should be a data logger that records when an automatic driving function is switched on or off and all other malfunctions and accidents happening in the vehicle as well.

---

[29] Law Web, "Important provisions of Information Technology Act 2000" *available at*: https://www.lawweb.in/2019/09/important-provisions-of-information.html (last visited November 7, 2023).
[30] Information Technology.
[31] Automated and Electric Vehicles Act, 2018.
[32] Automated and Electric Vehicles Act, 2018, s. 8(1)(a).
[33] Automated and Electric Vehicles Act, 2018, s. 2(1).
[34] Automated and Electric Vehicles Act, 2018, s. 2(3).
[35] Automated and Electric Vehicles Act, 2018, s. 2(4).
[36] Automated and Electric Vehicles Act, 2018, s. 9(1).
[37] Road Traffic Act (StVG), s. 1a(2).

   2.4 CHINA

China has issued several guidelines for SDVs. These standards mainly include the Guidelines on Road Testing of Intelligent Connected Vehicles (Trial)[42] and the Safety Technical Requirements for Road Testing of Intelligent Connected Vehicles (Trial)[43]. They declare an intelligent connected automobile to be a motor vehicle that can replace partially or wholly human driving through advanced sensors, controllers and actuators as well as other devices and is capable of perceiving traffic environment information, planning driving routes and executing longitudinal and lateral control. Additionally, these documents establish that during road testing, the individual conducting the test is responsible for any damage caused by the vehicle unless it can be proven that such damage was due to force majeure or an intentional act by a third party.[44] The tester must also obtain a testing permit; ensure that there's a qualified safety driver present in the vehicle who can take over at any time; install data recorder and remote monitoring system in the vehicle; and report all accidents or failures to authorities according to these documents.[45]

---

[38] Straßenverkehrsgesetz.
[39] Road Traffic Act (StVG), s. 7(1).
[40] Road Traffic Act (StVG), s. 7(3).
[41] "Self-driving car liability," *Wikipedia*, 2023.
[42] Shanghai Municipal People's Government, "Announcement of the Standing Committee of the 15th Shanghai Municipal People's Congress" *available at*:
https://www.shanghai.gov.cn/nw48050/20230317/49999fa1d26e49e3b887b5dbe0751490.html (last visited November 7, 2023).
[43] Conventus Law, "China - National Administrative Rules of Road Testing of Self-driving Vehicles Promulgated." *available at*: https://conventuslaw.com/report/china-national-administrative-rules-of-road/ (last visited November 7, 2023).
[44] China Daily, "Fresh guideline highlights autonomous vehicle tests" *available at*:
https://english.www.gov.cn/policies/policywatch/202211/03/content_WS6363176ec6d0a757729e23ac.html (last visited November 7, 2023).
[45] Iris Deng, "China steps up autonomous driving development with new guidelines on operating driverless vehicles for public transport" *available at*: https://www.scmp.com/tech/policy/article/3188314/china-steps-autonomous-driving-development-new-guidelines-operating (last visited November 7, 2023).

These are examples of how different nations have dealt with the matter of self-driving cars and their criminal accountability. However, there are many loopholes in this area. These countries clearly lack a harmonized approach when it comes to regulations. For example, there is no clarity on what constitutes a SDV or its level of autonomy. There is no consistency in terms of criteria to be used by persons trying to establish responsibility in cases of accidents. Dispute resolution mechanism is non existent, while at the same time, enforcing liability claims has become problematic. The cost of protecting data privacy and security appears inadequate. Equally, there are no incentives for promoting innovation and cooperation among stakeholders that would be effective in doing so. Therefore, more research into this issue, as well as better coordination between countries, needs to be undertaken urgently. Likewise, increased public awareness and education about the subject must go hand-in-hand with greater participation and consultation by concerned groups within society. Lastly, a balanced approach that equally takes into account the other aspects of SDV other than the legality is necessary in order to address any problem encountered during their operation process.

## 3. COMPARING LIABILITY CLAIMS IN AVs

This part draws comparisons involving liability claims on SDVs from different countries. It contains the way different courts or authorities dealt with some cases as well as the legal arguments that were made by parties to those cases. Also, it has implications for what could happen in the future and how regulations may be developed for autonomous vehicles going forward.

### 3.1 INDIA

India is one of the high-growth opportunity market for automobiles globally.[46] Alarmingly, it's also amongst the countries with highest road accident death toll. As per MoRTH[47], there were a total of 4,49,002 accidents in 2019. These mishaps resulted in 1,51,113 deaths and 4,51,361 injuries (approx.).[48] Most of these were caused because of human error; primarily consequential of overspeeding, drunk driving as well as breach of traffic rules, among others. Consequently, SDVs seem to have a possibility to decrease these collisions and boost highway safety in India.[49] However, at present, India does not have any specific legal framework or regulation that deals with autonomous vehicles explicitly.[50] This main law overseeing motorcars which, is Motor Vehicles Act, 1988 (MVA), neither recognizes nor defines what self-driving vehicles are. It does not provide their different levels of automation as well. The MVA also fails to discuss the matter about liability when it comes to SDVs. There is no provision relating to their manufacturers or developers too.[51] The only pertinent section in MVA is Section 146. This section makes third-party insurance obligatory. Nonetheless, this provision does not state how such liability will be determined in cases involving SDVs. The implication of this is that the policy must cover the vehicle for any liabilities that may arise from death or bodily injury to a person or damage to any property resulting from the use of it.[52]

---

[46] Invest India, "Invest in Indian Automobile Industry, Auto Sector Growth Trends" *available at*: https://www.investindia.gov.in/sector/automobile (last visited November 7, 2023).
[47] Ministry of Road Transport and Highways.
[48] Megha Sood, "India had most deaths in road accidents in 2019: Report" *Hindustan Times* (Mumbai, 25 October 2020).
[49] Team Ackodrive, "Causes of Road Accidents in India - Why Do Accidents Occur?" *available at*: https://ackodrive.com/traffic-rules/causes-of-road-accidents/ (last visited November 7, 2023).

As a result, there is a void in the Indian legislation concerning autonomous vehicles. Therefore, there will be no sufficient grounds to make an argument about liability claims of self-driving cars until proper laws are put in place. Consequently, any case that arises from such cases will need to consider the general principles of tort law or contract law applicable in India. For example, if this vehicle causes a road accident which leads to the death or injury of another person or damage to somebody else's property, the injured party may go ahead and file a lawsuit against the owner or even operator of such a motor vehicle on account of negligence and breach of duty by care.[53] The victim can also sue for product liability or defective design, for instance, the manufacturers or developers of this vehicle.[54] In respect to either eventuality, plaintiffs must establish a link between what was done by defendants and the pain caused as their burden within this context. Defendants can plead contributory negligence, assumption of risk and exculpatory excuses depending on the circumstances in order to avoid or mitigate liability for their actions. The defendant might also have other defences like force majeure et alia that he/she can employ so as not to be held accountable at all or lessen her/his guilt before court proceedings.[55]

---

[50] Sharath Kumar Nair, "Self-driving cars to become a major challenge for legal systems" *available at*: https://analyticsindiamag.com/self-driving-cars-to-become-a-major-challenge-for-legal-systems/ (last visited September 10, 2023).
[51] Tejas Sateesha Hinder and Ritik Kumar Rath, "Self-Driving Cars and India: A Call for Inclusivity under the Indian Legal Position" *available at*: https://lawreview.nmims.edu/self-driving-cars-and-india-a-call-for-inclusivity-under-the-indian-legal-position/ (last visited November 7, 2023).
[52] Essenese Obhan and Shourya Paul, "Driving into the Future: Regulating Autonomous Vehicles" *available at*: https://www.mondaq.com/india/patent/1239240/driving-into-the-future-regulating-autonomous-vehicles (last visited November 7, 2023).
[53] Om Shivam, "Analysis of laws regulating self-driving cars" *available at*: https://blog.ipleaders.in/regulation-of-self-driving-cars/ (last visited November 7, 2023).
[54] "Self-driving car liability," *Wikipedia*, 2023.
[55] Patrick H. Reilly and Elsa M. Bullard, "5 Defenses for Autonomous Vehicles Litigation" *Faegre Drinker* 1–8 (2018).

## 3.2 USA

Among other countries, USA is at the forefront of designing and testing autonomous vehicles[56], with several companies such as Tesla, Waymo, Uber, etc, operating or experimenting with them on public roads.[57] Nevertheless, there lacks a uniform federal legal framework or regulation for self-driving cars.[58] Instead, there is a wide variety of regulations & state laws that are not equal in scope and content.[59] According to NCSL[60] data as of March 2020, 29 states and the District of Columbia have enacted legislation related to autonomous vehicles and executive orders have been issued in 11 states on the same. These statutes govern different aspects of autonomous vehicle technology, including definitions, testing, deployment, licensing, registration, insurance, liability, safety, data, etc.[61]

Nonetheless, these rules differ and cannot be comprehensive enough to cover all the problems associated with autonomous vehicles.[62] In addition, they can come into conflict with and be pre-empted by other federal statutes that are applicable to motor vehicles generally speaking.[63] NHTSA[64] is responsible for setting regulations concerning automotive safety standards across the country. It also ensures safety standards in cases of the equipment used.[65] NHTSA has issued some guidelines on SDVs, which include, Automated Driving Systems 2.0[66] & 3.0[67]. However, these documents have no force or effect; and they do not expressly address the liability issue surrounding autonomous driving in clear language.[68]

Hence, similar to India, a self-driving vehicle liability claim in the USA will be decided on an individual case basis through courtrooms using general principles of tort law or contract law.[69] In contrast to India, there have been several reported cases or incidents involving self-driving vehicles in the United States that have sparked off liability concerns.[70] A few such accounts are as follows:

---

[56] Johannes Deichmann et al., *Autonomous Driving's Future: Convenient and Connected* 1–12 (McKinsey & Company, 7 November 2023).
[57] Sunny Betz, "28 Self-Driving Car Companies You Should Know" *available at*: https://builtin.com/transportation-tech/self-driving-car-companies (last visited November 7, 2023).
[58] Eric Stauffer and Brad Larson, "Which states allow self-driving cars? (2023)" *available at*: https://www.autoinsurance.org/which-states-allow-automated-vehicles-to-drive-on-the-road/ (last visited November 7, 2023).
[59] "Regulation of Self-Driving Cars," *Wikipedia*, 2023.
[60] National Conference of State Legislatures.
[61] National Conference of State Legislatures, *Autonomous Vehicles | Self-Driving Vehicles Enacted Legislation*, 18 February 2020.
[62] Justin Banner, "Are Autonomous Self-Driving Vehicles Legal in My State?" *available at*: https://www.motortrend.com/features/state-laws-autonomous-self-driving-driverless-cars-vehicles-legal/ (last visited November 7, 2023).
[63] Jones Day, "Autonomous Vehicles: Legal and Regulatory Developments in the United States" *Jones Day White Paper* (2021).
[64] National Highway Traffic Safety Administration.
[65] NHTSA, *Federal Automated Vehicles Policy* 1–116 (U.S. Department of Transportation, September 2016).
[66] NHTSA, *Automated Driving Systems 2.0: A Vision for Safety* 1–36 (U.S. Department of Transportation, September 2016).
[67] NHTSA, *Preparing for the Future of Transportation: Automated Vehicle 3.0* 1–80 (U.S. Department of Transportation, October 2018).
[68] Jon Eberst, "Self-Driving Cars and Liability" *available at*: https://eberstlaw.com/2020/02/06/self-driving-cars-and-liability/ (last visited November 7, 2023).

### 3.2.1    AUTOPILOT CRASHES

Tesla is a top manufacturer of EVs[71] and SDVs globally.[72] Its Autopilot mode is an attribute for its cars to steer, accelerate and brake. It can also change lanes and park itself under certain circumstances.[73] But still, the company cautions its customers that Autopilot is not fully autonomous. The drivers must pay attention and be prepared to take control at any time.[74] But even with this warning, there have been several accidents involving Tesla cars operating in this very mode. Tesla has been charged with negligence or product liability on its drivers because of some crashes that took place. These mishaps result in fatal injuries to drivers or passengers or other road users.[75]

In 2016, after a fatal crash into a tractor-trailer, Joshua Brown got killed. His Tesla Model S was in auto-pilot mode. The incident took place in the month of May on a highway in Florida.[76] The NHTSA probed the accident and found that Brown ignored seven visual warnings from the Autopilot system before colliding with it despite six voice warnings from the same system.[77] Another NHTSA's investigation also showed that the autopilot system could not detect the white tractor-trailer due to its light colour against a bright sky. The Autopilot system was not found to have any fault by NHTSA. Hence, no action against Tesla.[78] Nevertheless, Brown's family filed a lawsuit against Tesla in the year 2018. They claimed for the wrongful death and product liability. The suit claimed that the company was at fault for negligently selling the Autopilot system. They claimed this as it failed to warn Brown of its limitations. Additionally, the suit alleged that the company had misrepresented the Autopilot as reliable while it was actually dangerous.[79] Consequently, there was a confidential settlement of this matter in 2019.

---

[69] Steven D. Jansma, "Autonomous vehicles: The legal landscape in the US" *Norton Rose Fulbright* 1–26 (2016).
[70] Tom Krisher, "US report: Nearly 400 crashes of automated tech vehicles" *AP News*, 16 June 2022.
[71] Electric Vehicles.
[72] Tesla, "Autopilot and Full Self-Driving Capability" *available at*: https://www.tesla.com/support/autopilot (last visited November 7, 2023).
[73] "Tesla Autopilot," *Wikipedia*, 2023.
[74] Sky UK, "Tesla full self-driving software 'may do the wrong thing at the worst time' company warns" *Sky News*, 12 July 2021.
[75] "List of lawsuits involving Tesla, Inc.," *Wikipedia*, 2023.
[76] Danny Yadron and Dan Tynan, "Tesla driver dies in first fatal crash while using autopilot mode" *The Guardian*, 30 June 2016.
[77] David Shepardson, "Tesla driver in fatal 'Autopilot' crash got numerous warnings: U.S. government" *Reuters*, 19 June 2017.
[78] Jack Stewart, "After Investigating Tesla's Deadly Autopilot Crash, Feds Say Hooray for Self-Driving" *Wired*, 2023.

In 2019, Walter Huang was involved in an unfortunate incident. The crash happened on Highway-101 in California. His Tesla Model X went off course, colliding with a concrete barrier, resulting in his demise. NTSB[80] conducted an investigation into the accident and concluded that before the collision occurred, there had been several warnings to Huang from the system to put his hands back on the steering wheel. In addition, NTSB also found out that Autopilot misread lane markers and directed his vehicle to hit a concrete divider. When considering road conditions under which Autopilot mode should be used as well as preventing driver distraction, NTSB lambasted Tesla for failing to do so.[81] His family has sued Tesla in court since they believed this system led to his death through negligence and faulty product liability. They claimed that this was due to negligent designing, testing, marketing and selling of the system. Furthermore, it was alleged that Tesla did not properly warn Huang about its dangers. It represented the device falsely as safe, even though it was dangerous.[82] The case remains pending at court.[83]

Jeremy Banner's fatality happened in Florida in December 2019. The Tesla Model 3 was being driven by Jeremy Banner himself. It was in Autopilot mode when it collided with a tractor-trailer along Highway I-75 in Florida. In response to the accident, NTSB looked into the crash, establishing Banner had turned on the Autopilot system just seconds before running his vehicle into a semi-truck trailer and he never tried to avoid the accident.[84] The board also noted that the system could not detect the tractor-trailer because of its colour combined with the lighting conditions.[85] In the year 2020, an action for wrongful death and product liability was filed against Tesla by members of Jeremy Banner's family. The lawsuit claimed that Tesla failed in its duty of care while designing, testing, marketing and selling the Autopilot system; as well as failed to alert him about its limitations or dangers as they were known to him at the time or what could be reasonably understood based on his experience with other self-driving cars. The complaint further alleges that Tesla misrepresented that it was safe to operate the Autopilot so long as drivers maintained some degree of attention.[86] This case is still pending before court.[87]

---

[79] Amy Martyn, "Lawsuit charges Tesla of misleading consumers about safety of its Autopilot feature" *Consumer Affairs*, 28 February 2018.
[80] National Transportation Safety Board.
[81] BBC News, "Tesla Autopilot crash driver 'was playing video game'" *BBC News*, 26 February 2020.
[82] Richard Lawler, "Tesla sued over fatal 2018 Model X crash with Autopilot engaged" *available at*: https://www.engadget.com/2019-05-01-tesla-autopilot-lawsuit-model-x.html (last visited November 7, 2023).
[83] "List of lawsuits involving Tesla, Inc.," *Wikipedia*, 2023.
[84] Andrew J. Hawkins, "Tesla didn't fix an Autopilot problem for three years, and now another person is dead" *available at*: https://www.theverge.com/2019/5/17/18629214/tesla-autopilot-crash-death-josh-brown-jeremy-banner (last visited November 7, 2023).
[85] Timothy B. Lee, "'I was just shaking'—new documents reveal details of fatal Tesla crash" *available at*: https://arstechnica.com/cars/2020/02/i-was-just-shaking-new-documents-reveal-details-of-fatal-tesla-crash/ (last visited November 7, 2023).

### 3.2.2 UBER SDV FATALITY CASE

Uber is another company that has been testing AVs on public roads in the USA. It's developed SDVs are fitted with detectors, video cameras, radar sensors, lidars as well as software to allow them manoeuvre through traffic. Nonetheless, actual human security drivers are still hired to remain behind the steering wheel, watching over the vehicle, ready to intervene if need be.[88]

Towards the end of March 2018, one of its AVs hit a pedestrian named Elaine Herzberg crossing a street at night in Tempe, Arizona. This was the first known death involving an autonomous vehicle in America, which led to public outrage. This sparked debate about the safety and regulation of self-driving cars. These are some findings by NTSB concerning:

i. The AV was operated in auto mode but had a human driver for safety purposes when this occurred.[89]

ii. The AV detected Herzberg 5.6 seconds before impact, but failed to accurately identify her or predict her path correctly.

iii. It did not brake or steer to avoid Herzberg because Uber had disabled its self-driving system's emergency braking feature so as to minimize potential clashes with human intervention.[90]

iv. The person responsible for monitoring the vehicle was distracted by her own personal mobile phone and failed to keep track of either the road or the performance of the car until half a second before impact occurred.[91]

v. The human safety driver could have avoided the crash if she had been attentive and reacted appropriately.[92]

---

[86] Isobel Asher Hamilton, "'We cannot have technology and sales take over safety': Tesla is being sued again for a deadly Autopilot crash" *available at*: https://www.businessinsider.in/we-cannot-have-technology-and-sales-take-over-safety-tesla-is-being-sued-again-for-a-deadly-autopilot-crash/articleshow/70499578.cms (last visited November 7, 2023).
[87] HT Auto Desk, "Tesla CEO Elon Musk spared from testifying in Autopilot crash lawsuit" *Hindustan Times*, 18 March 2022.
[88] Leif Johnson and Michelle Fitzsimmons, "Uber self-driving cars: everything you need to know" *available at*: https://www.techradar.com/news/uber-self-driving-cars (last visited November 7, 2023).
[89] Sam Levin and Julia Carrie Wong, "Self-driving Uber kills Arizona woman in first fatal crash involving pedestrian" *The Guardian*, 19 March 2018.
[90] Mark Harris, "NTSB Investigation Into Deadly Uber Self-Driving Car Crash Reveals Lax Attitude Toward Safety" *available at*: https://spectrum.ieee.org/ntsb-investigation-into-deadly-uber-selfdriving-car-crash-reveals-lax-attitude-toward-safety (last visited November 7, 2023).
[91] Matt McFarland, "Feds blame distracted test driver in Uber self-driving car death" *available at*: https://edition.cnn.com/2019/11/19/tech/uber-crash-ntsb/index.html (last visited November 7, 2023).

Uber's insufficient safety culture was found by NTSB as most likely cause of the accident, which also led to its failure to address testing risks of autonomous vehicles on public roads. The NTSB also blamed the human safety driver for her inattention and failure to intervene, as well as Herzberg for crossing outside a crosswalk at night. In April 2018, Herzberg's family reached a confidential settlement with Uber for an undisclosed amount of money.[93] The crash also led to criminal and civil actions against Uber or its safety driver. The proceedings took place in the month of March of 2019. Yavapai County Attorney's Office announced no criminal charges for the crash. This was because of lack of evidence, necessary to prove criminal liability.[94]

In August 2019, Maricopa County Attorney's Office charged Rafaela Vasquez with negligent homicide leading to the death of Herzberg. She was the human safety driver. Vasquez denied the charges and is expecting a trial.[95] This case demonstrates how complex and uncertain it is to determine and allocate responsibility for autonomous vehicles, in particular with human components that can be distracted, intervened in or make mistakes. It also signifies the need for coherence clarity on standards and rules guiding the trial of self-driving vehicles on public roads.

### 3.3 UK

The United Kingdom is an actively involved country in developments towards testing of automated cars, including projects like UK Autodrive, GATEway, Venturer, among others.[96] Additionally, the British government has announced its backing for autonomous vehicles aiming at making UK as global leader in this field, hence, fostering safe environment and encouraging invention and usage of these machines.[97]

The UK government passed the Automated and Electric Vehicles Act (AEVA) in 2018, the first-ever legislation globally to focus on self-driving vehicle liability.[98] In order for a vehicle to be regarded as a self-driving one according to AEVA, there must be cases where it can drive safely with no human control or observation. Besides, AEVA also provides that there should be a list of vehicles which are self-driving as specified by the Secretary of State.

---

[92] Rory Cellan-Jones, "Uber's self-driving operator charged over fatal crash" *BBC News*, 16 September 2020.
[93] Reuters, "Uber settles with family of woman killed by self-driving car" *The Guardian*, 29 March 2018.
[94] 12News, "Uber driver charged in self-driving crash that left woman dead in Tempe in 2018" *12News*, 15 September 2020.
[95] Corina Vanek, "Arizona driver in fatal autonomous Uber crash in 2018 pleads guilty, sentenced to probation" *The Arizona Republic*, 28 July 2023.
[96] Iain Forbes, *Connected and Autonomous Vehicles in the UK* 1–26 (Center of Connected & Autonomous Vehicles, July 2016).
[97] Rachel Maclean, "Government paves the way for self-driving vehicles on UK roads" *available at*: https://www.gov.uk/government/news/government-paves-the-way-for-self-driving-vehicles-on-uk-roads (last visited November 7, 2023).
[98] Matthew Channon, "Automated and Electric Vehicles Act 2018: An Evaluation in light of Proactive Law and Regulatory Disconnect," 10 *European Journal of Law and Technology* 1–36 (2019).

As per AEVA:

1. Regardless of whether a vehicle is insured or uninsured; if it gets into an accident while driving itself, the blame will go to its owner or the insurance company.
2. If the cause of the incident is as follows, then such damage or injury may be compensated for by the insurer or owner of a SDV from the person responsible for the car:
   i. The vehicle was made to move on its own in circumstances where it would have been inappropriate.
   ii. The individual supervising the vehicle failing to take reasonable measures that could prevent an accident.
   iii. The vehicles software or hardware was tampered with or otherwise altered in such a way that changed its operation.
3. An insurance company will not be held answerable for any wrongdoing caused by an autonomous automobile under their cover if:
   i. Another person's actions that result in destruction and injury were intended.
   ii. Terrorism and war are involved.
   iii. An unavoidable occurrence due to natural causes.

AEVA provides that:

1. A driver of a self-driven car cannot be charged with any criminal offence arising from its act of driving itself except he or she disrupted its working order or commanded it to do so when it was unnecessary.
2. The maker or developer of a driverless vehicle cannot be sued for civil damages brought about by such cars unless they supplied defective goods, broke their promise or warranty agreement.

In the UK, AEVA is intended to make it easy for people to understand who is liable for an accident involving self-driving vehicles. Also, the legislation means that insurers and owners can have confidence in a system that provides adequate coverage.[99]

### 3.4 GERMANY

Several companies, including BMW, Mercedes-Benz, Volkswagen, etc. are heavily investing in this area, making Germany the centre of autonomous driving development and testing.[100]

---

[99] Automated and Electric Vehicles Act, 2018.

Germany is also known for its automotive industry prowess and engineering knowledge base as well as the rigid safety measures plus regulations set by the country.[101] Based on this reference point, Germany was one of the first countries globally to legislate on autonomous cars when it amended its Road Traffic Act (Straßenverkehrsgesetz).[102] The amendment categorizes highly automated driving systems as those which can perform every task related to driving within certain parameters without needing a human being's input. Furthermore, according to the amendment, driver assistance systems are able to aid or support humans in some driving activities, but they need somebody else to watch out for them.[103]

The amendment provides that:

1. A driver of a highly automated driving system shall not be held responsible for any harm or damage caused by the system while it is within its stated use case, unless he or she inhibited its operation or made it operate other than its specific use case.
2. The driver of such a highly automated driving system will hold any damages inflicted by the system during operation outside of its specified use case; they can only be absolved from blame if they assumed control of the vehicle immediately or the accident was unavoidable.
3. When in operation, the person who operates should bear all blame for the system's defects resulting in injuries and damages, except where there was no negligence on his part or where an accident is not avoidable.
4. Unless he/she delivered a faulty product, breached a contract/warranty, supplied a defective product, neither civil liability will attach to the maker/developer of such a highly-automated vehicle nor to that of any driver-assistance-systems during operation.

The amendment also provides that:

1. The driver of a highly automated driving system must keep awake and be ready to take over the car if not ordered by the system otherwise.

---

[100] Christy Gren, "Germany takes a pioneering role in self-driving cars development" *available at*: https://www.industryleadersmagazine.com/germany-takes-a-pioneering-role-in-self-driving-cars-development/ (last visited November 7, 2023).
[101] BMDV, "Germany will be the world leader in autonomous driving" *available at*: https://bmdv.bund.de/SharedDocs/EN/Articles/DG/act-on-autonomous-driving.html (last visited November 7, 2023).
[102] Dr. Gerd Leutner and Dr. Martin Eichholz, "Autonomous vehicles law and regulation in Germany" *available at*: https://cms.law/en/int/expert-guides/cms-expert-guide-to-autonomous-vehicles-avs/germany (last visited November 7, 2023).
[103] Jenny Gesley, "Germany: Road Traffic Act Amendment Allows Driverless Vehicles on Public Roads" *available at*: https://www.loc.gov/item/global-legal-monitor/2021-08-09/germany-road-traffic-act-amendment-allows-driverless-vehicles-on-public-roads/ (last visited November 7, 2023).

2. A driver of a highly automated driving system has to follow the road rules applicable to humans unless otherwise instructed by the system.
3. Furthermore, this new provision requires that a driver using a fully automated driving system must capture and store all data relating to its operation and make it accessible to authorities during investigations of accidents or other incidents.
4. Moreover, according to these new provisions, a manufacturer or developer of such an automated vehicle will have to ensure that it complies with technical and safety requirements prescribed by the Federal Motor Transport Authority (Kraftfahrt-Bundesamt) and obtain its consent before releasing it for use on public roads.

The amendment aims at clarifying liability issues in relation to AVs in Germany. Additionally, it seeks to give enough protection, both for drivers as well as manufacturers so that they can develop and adopt them.[104]

### 3.5 CHINA

Several companies in China (including Baidu, Alibaba, Tencent, etc.) are racing to develop and test AVs[105]. Besides, the country has a huge market potential plus the demand for SDVs due to population growth, urbanization, traffic congestion, pollution, among others.[106]

Nonetheless, there is currently no dedicated legal framework or regulation for the operation of self-driving vehicles in China.[107] The existing laws affecting motor vehicles generally fall short of adequately addressing SDVs. For example, the Road Traffic Safety Law (2003) mandates that all motor vehicles should be driven by a person who has the requisite driving skills and knowledge.[108] Another, Tort Liability Law (2009) stipulates that the driver of a vehicle is liable for any injury or damage resulting from his/her fault or negligence.[109] These laws do not recognize or define SDVs or their levels of automation. They also do not address

---

[104] Christian M. Theissen, "The new German bill on automated vehicles – and the resulting liability changes," in M. Bargende, H.-C. Reuss, *et al.* (eds.), *18. Internationales Stuttgarter Symposium* 1295–301 (Springer Fachmedien, Wiesbaden, 2018).
[105] Luca Pizzuto et al., "The market for China autonomous vehicles" *available at*: https://www.mckinsey.com/industries/automotive-and-assembly/our-insights/how-china-will-help-fuel-the-revolution-in-autonomous-vehicles (last visited November 7, 2023).
[106] Eduardo Jaramillo, "Top 10 Chinese autonomous driving companies to watch" *available at*: https://thechinaproject.com/2023/02/23/ten-chinese-autonomous-driving-companies-to-watch/ (last visited November 7, 2023).
[107] Mark Schaub and Atticus Zhao, "China's Legislation on Autonomous Cars Rolls out" *available at*: https://www.chinalawinsight.com/2021/04/articles/corporate-ma/chinas-legislation-on-autonomous-cars-rolls-out/ (last visited November 7, 2023).
[108] "Road Traffic Safety Law of the People's Republic of China," *Wikipedia*, 2023.
[109] Ministry of Commerce People's Republic of China, "Tort Law of the People's Republic of China" *available at*: http://english.mofcom.gov.cn/article/policyrelease/Businessregulations/201312/20131200432451.shtml (last visited November 7, 2023).

the issue of liability for self-driving cars or their manufacturers.[110] Therefore, similar to India[111] or USA[112], any litigation concerning auto-robotics will have to be decided by Chinese courts on an individual basis under tort law principles or contracts law.[113] There have been no reported cases or incidents involving self-driving vehicles in China that have raised or triggered liability issues. Therefore, it is difficult to predict how courts will approach and resolve these issues.[114] But some scholars and experts suggested some ideas and recommendations on establishing a legal framework or regulation for self-driving vehicles in China:

This means the developers of autonomous vehicles should be held strictly liable whenever their products cause injuries or damages, whether they are at fault for these defects or not.[115] This kind of system would enable victims to obtain fair compensation while motivating manufacturers and developers to enhance safety and quality features of self-driving cars. The country's leaders must establish a mandatory insurance policy which accommodates for all mishaps caused by the manufacturers.[116] As concerns over the question of responsibility continue to generate discourse among Chinese academicians, some suggestions have been advanced on this matter. Nevertheless, those views cannot be regarded as official interpretations because they may not necessarily reflect what happens now or later with regards to pronouncements made by courts or other bodies within China.

## 4. KEY FINDINGS FROM THE COMPARATIVE STUDY

There are complex legal challenges in different areas of law that affect autonomous vehicles. Global legal framework for driverless cars is not uniform and this highlights the need to tackle issues of responsibility, safety as well as interaction between man and machine together. The comparisons drawn among India, the USA, UK, Germany and China shows there is an immediate need for an all-encompassing and adaptable legal regime that can embrace or deal with the complexities involved in this transformational technology. The rationale for comparing these countries lies in their diverse legal frameworks. These countries were chosen for their technological advancements and contrasting regulatory approaches to liability in AI-enabled autonomous vehicles. This paper has probed into the intricate legal landscape surrounding autonomous vehicles across various jurisdictions. Using existing laws, cases and regulatory trends as a source of data, this research highlights some key insights and recommendations.

---

[110] "Civil and Commercial Laws," *available at*: http://www.npc.gov.cn/zgrdw/englishnpc/Law/2011-02/16/content_1620761.htm (last visited November 7, 2023).
[111] Disha Patwa, "Autonomous vehicles: The question of liability" *available at*: https://lawbeat.in/articles/autonomous-vehicles-question-liability (last visited November 7, 2023).
[112] Steven D. Jansma, "Autonomous vehicles: The legal landscape in the US" *Norton Rose Fulbright* 1–26 (2016).
[113] Rita Liao, "Real driverless cars are now legal in Shenzhen, China's tech hub" *available at*: https://techcrunch.com/2022/07/25/real-driverless-cars-legal-in-chinas-shenzhen/ (last visited November 7, 2023).
[114] Mark Schaub and Atticus Zhao, "China's Legislation on Autonomous Cars Rolls out" *available at*: https://www.chinalawinsight.com/2021/04/articles/corporate-ma/chinas-legislation-on-autonomous-cars-rolls-out/ (last visited November 7, 2023).
[115] Jiaxin Liu et al., "Road Traffic Law Adaptive Decision-making for Self-Driving Vehicles" *2022 IEEE 25th International Conference on Intelligent Transportation Systems (ITSC)*, 2022.
[116] Yineng Xiao and Zhao Liu, "Accident Liability Determination of Autonomous Driving Systems Based on Artificial Intelligence Technology and Its Impact on Public Mental Health," 2022 *Journal of Environmental and Public Health* e2671968 (2022).

The fact that there is no global uniform framework that governs the use of autonomous vehicles reveals how intricate it is to regulate this type of technology. Different countries have different legal, cultural and technological factors that they consider in forming their responses to AV liability. The analysis illustrates the urgency for comprehensive and adaptable legal systems that can tackle the complex issues arising from AV rollout. The examples demonstrate how complex it can be to attribute blame for an accident when both driver and car are involved in the decision-making process; for example, Tesla's Autopilot crashes in America or Uber's lethal car crash in Arizona demonstrate this. The UK has been proactive by enacting specific legislations for liability caused by AVs compared to India, which has not yet regulated on this matter. Inadequate frameworks create significant problems for various stakeholders, including manufacturers, insurers and consumers alike. It underlines the importance of worldwide harmonization of legal norms in order to boost innovation, ensure responsibility and reduce risks within increasingly automated transport systems. In the process of establishing international guidelines and standards for AVs, it is important that they be designed to prioritize safety, equity and adaptability for different legal systems and cultures. Thus, these findings result in various recommendations:

i. The policy makers should focus on developing and implementing clear, flexible and technology-neutral regulatory frameworks that provide certainty and accountability to all stakeholders.

ii. More sharing of knowledge and collaboration between countries' industry players as well as regulatory institutions are vital to harmonizing the AV rules globally.

iii. Ongoing researches should be done together with dialogue aimed at addressing emerging legal challenges, especially those associated with AVs such as data privacy, cyber security, among others.

| Country | Legal Framework | Liability Model | Relevant Provisions/Guidelines | Key Differences |
|---------|-----------------|-----------------|--------------------------------|-----------------|
| **India** | Motor Vehicles Act, 1988 | Tort Law | Section 146 (Third-party insurance) | Lack of specific regulations |
| **USA** | State laws, NHTSA | Tort Law, Product Liability | NHTSA guidelines, State legislation | Variation in state laws |
| **UK** | Automated and Electric Vehicles Act, 2018 | AEVA provisions | Liability for self-driving incidents, Insurer responsibility | First legislation globally |
| **Germany** | Road Traffic Act (StVG) | Direct Liability | Liability for defects, Compliance with safety standards | Clear regulations on AV operation |
| **China** | Various guidelines | Suggestions for liability laws | General laws applicable, Proposed liability framework | Lack of dedicated regulations |

Table 1.1 Legal Models in Selected Countries for Autonomous Vehicles

Consequently, it is very necessary for legal frameworks to move with the pace of technological innovation in AVs so that safe autonomous vehicle deployment may transpire concurrently. By upholding safety fundamentals yet embracing innovation, policymakers can effectively utilize AV's transformative potential by creating a more efficient transportation system which is sustainable, fairer and equitable for future generations. As autonomous vehicles continue to advance, global harmonization of legal standards should be pursued so as to encourage creativity, ensure responsibility and reduce risks within an increasingly automated transport system. As an outcome of this comparative study, it is found that there is a pressing need for globally agreed upon legal standards to encourage further technological advancements.